# A Quantitative Framework for Establishing Low-risk Interdistrict Travel Corridors during COVID-19

**Raviraj Dave**
Discipline of Civil Engineering
Indian Institute of Technology, Gandhinagar, Gujarat, 382355, India
Email: dave_raviraj@iitgn.ac.in

**Tushar Choudhari**
Department of Civil Engineering
Indian Institute of Technology Bombay, Powai, Mumbai, 400076, India
Email: tp.choudhari@gmail.com

**Udit Bhatia, Ph.D.**
Discipline of Civil Engineering
Indian Institute of Technology, Gandhinagar, Gujarat, 382355, India
Email: bhatia.u@iitgn.ac.in

**Avijit Maji, Dr. Engg., P.E., PTOE**
Department of Civil Engineering
Indian Institute of Technology Bombay, Powai, Mumbai, 400076, India
Email: avijit.maji@gmail.com

Word Count: 6254 words + 1 Table (250 words) = 6554 words

*Submitted: October 31, 2020*

*Raviraj Dave, Tushar Choudhari, Udit Bhatia and Avijit Maji*

**ABSTRACT**

Aspirations to slow down the spread of Novel Coronavirus (SARS-CoV2) resulted in unprecedented restrictions on personal and work-related travels in various nations across the globe. As a consequence, economic activities within and across the countries were almost halted. As restrictions loosen and cities start to resume public and private transport to revamp the economy, it becomes critical to assess the commuters' travel-related risk in light of the ongoing pandemic. We develop a generalizable quantitative framework to evaluate the commute-related risk arising from inter-district and intra-district travels by combining Nonparametric Data Envelopment analysis for vulnerability assessment with transportation network analysis. We demonstrate the application of the proposed model for establishing travel corridors or travel bubbles within and across Gujarat and Maharashtra, two Indian states that have reported many SARS-CoV2 cases since early April 2020. Our findings suggest that establishing the travel bubble between a pair of districts solely based on the health vulnerability indices of origin-destination discards the en-route travel risks due to prevalent pandemic, hence underestimating the threat. For example, while the resultant of social and health vulnerabilities of Narmada and Vadodara's districts is relatively moderate, the en-route travel risk exacerbates the overall travel risk. Our study provides actionable insights to users into choosing the alternate path with the least risk and can inform political decisions to establish low-risk travel corridors within and across the states while accounting for social and health vulnerabilities in addition to transit-time related risks.

*Raviraj Dave, Tushar Choudhari, Udit Bhatia and Avijit Maji*

**INTRODUCTION**

Since the first reported case of Novel Coronavirus (SARS-CoV2) in Wuhan, China, in mid-December 2019, 215 Countries and Territories have reported around 43 Million confirmed cases with more than 1.1 million fatalities as of October 25, 2020 (*1*, *2*). Two low-income countries that suffer the major impact of the spread in terms of number of reported cases of SARS-CoV2 include Brazil and India (*3*). While Brazil has reported nearly 5.4 Million SARS-CoV2 cases with more than 150 thousand deaths, India has reported 7.86 Million confirmed instances with more than 119 thousand deaths, as of October 25, 2020 (*1*).

For the first time in history, on March 24, 2020, a nationwide lockdown was imposed in India to contain the spread of SARS-CoV2, limiting the movement of 1.3 billion population (*4*, *5*). Policymakers worldwide have implemented massive travel restrictions and quarantining policies to mitigate the outbreak and reduce the stress on already reeling healthcare and emergency facilities, especially in developing countries and vulnerable communities (*6–9*). While the outbreak of deadly SARS-CoV2 took a toll on the individuals' lives and wellbeing (*10*), nationwide lockdowns severely impacted the economic activities within and across the country. The manufacturing, supply-chain, logistical activities, and retail activities have been severely crippled during the strict lockdown (*11–13*). Subsequently, to bring the impacted economy back on track, intrastate travel was permitted in a phased manner outside containment zones from the beginning of June 2020. Domestic flights have been allowed subject to the government's guidelines to ensure the passengers' safe travel amidst the pandemic. Besides, unrestricted vehicular movement within and across the states was permitted beginning late July-early August (*14*).

With unrestricted travel permitted, regions that reported fewer cases during the initial phase of disease spread started reporting a significant number of cases attributed to the increased population mobility from regions (*15*). While the control measures such as symptoms-based screening protocols were enacted at various airports and checkpoints, it is not sufficient to contain the pandemic spread as the majority of such cases arrive during an incubation period that is usually asymptotic (*16*). Moreover, following symptom-based surveillance protocols are challenging to enforce for road and rail transportation given multiple entries and exit points on the travel routes (*17*).

The travel bubble concept has emerged to address the tradeoffs between the attempts to control the disease spread by restricting human mobility and attempts to revive the economy by establishing travel regions across the regions (*18*). Travel bubbles, or travel corridors, are regional pacts between states that lower the travel barriers across the regions (*19*). While establishing travel corridors to restart commercial passenger services, trade, and economic activities are mutually beneficial (*20*), identifying travel corridors among regions solely based on case prevalence could result in underestimation of the risk.

We argue that identifying the interregional travel corridors needs to account for social and health vulnerabilities of the regions being linked and account for the possibilities of en-route exposure to the contagious diseases during the transit. The concept of establishing a travel corridor is based on the fact that regions being connected have comparable demography and socio-economic conditions. Moreover, the chances of importation of new caseload should be minimal to avoid further stresses on healthcare systems and emergency services. Hence, it is imperative to assess the social and health vulnerability indices and overall infection rates in the regions. While assessing social vulnerability on SaRS-CoV2 has gained significant attention at the regional and county-scale (*21*, *22*, *22*, *23*), possibilities of en-route risk to infectious diseases are seldom analyzed. Analysis of en-route exposure and risk becomes particularly crucial in the context of developing nations, including India, where road and rail constitute the most common mode of public transportation for intrastate and interstate travel (*24*). In mass public transit, adherence to physical distancing is looming in densely populated regions. While there is emerging evidence that the use of facemasks in closed settings can significantly reduce contagion risk (*25*), enforcing compliance is challenging. Hence, there is a growing concern that public transit can be viewed as unhealthy in the post-pandemic world. This belief, in turn, may translate behavioral changes of passengers and alter the choices from public mode to privately owned modes (*26*). In long-distance travel via road, drivers often develop motorist fatigue and use road-side parking or rest areas to alleviate the tiredness (*27*). Also, long-haul public transportation facilities have multiple stops en-route due to obligatory stops and to enable boarding and deboarding of the passengers (*28*). Thus, there is an increased risk of en-route exposure to highly prevalent and contagious diseases, including SARS-CoV2, from the public and private modes of transportation.

In the first-of-its-kind study, we develop a generalizable quantitative framework to identify low-risk inter-district travel corridors during the prevalence of SARS-CoV2. We evaluate the commute-related risk arising from inter-district and intra-district travels by combining Nonparametric Data Envelopment analysis (*29–32*) for social and health vulnerability assessment with the underlying transportation networks (*33, 34*). We demonstrate the application of the proposed model for identifying travel corridors within and across Gujarat and Maharashtra, the two Indian states that have reported a significant share of SARS-CoV2 cases since early April 2020. We further study the proposed approach's effectiveness by constructing the n-walk matrix between a pair of districts to identify all possible rerouting options and evaluating the least vulnerable route between the travel corridor's origin and destination nodes. Note that the clinical characteristics associated with the SARS-CoV2 (e.g., transmissibility, recovery, and mortality rates), varying travel speeds on various road segments within the states, congestions during peak-traffic hours, and administrative restrictions to travel through containment zone can impact the disease trajectory and hence the en-route risk of exposure in the real-time. Such effects are not included within the scope of this research.

## METHODS

### Data

We consider the neighboring states of Gujarat and Maharashtra, India, as the study area to demonstrate the application of the proposed framework. Located on the western coast of India, Gujarat and Maharashtra are one of the fastest-growing and leading industrialized states in India (*35*). With a combined population of 170.7 Million, Maharashtra and Gujarat are the $2^{nd}$ and $9^{th}$ most populous states in India and are home to the densely populated urban centers, including Mumbai and Ahmedabad (*36*). Given the high population densities, availability of opportunities, and continued financial activities increased the risk of transmission in both the states, Maharashtra has reported the highest number of cumulative cases as of October 25, 2020, whereas Gujarat reported the highest mortality rates in the initial phases of lockdown **(Figure 1)** (*37, 38*). Maharashtra and Gujarat rank among the top states in terms of manufacturing emergence and contribute nearly 50% to India's total export (*35*). Given the geographical proximity and high volumes of inter and intrastate transit volume, we analyze the inter-district travel in the 59 districts of the two states.

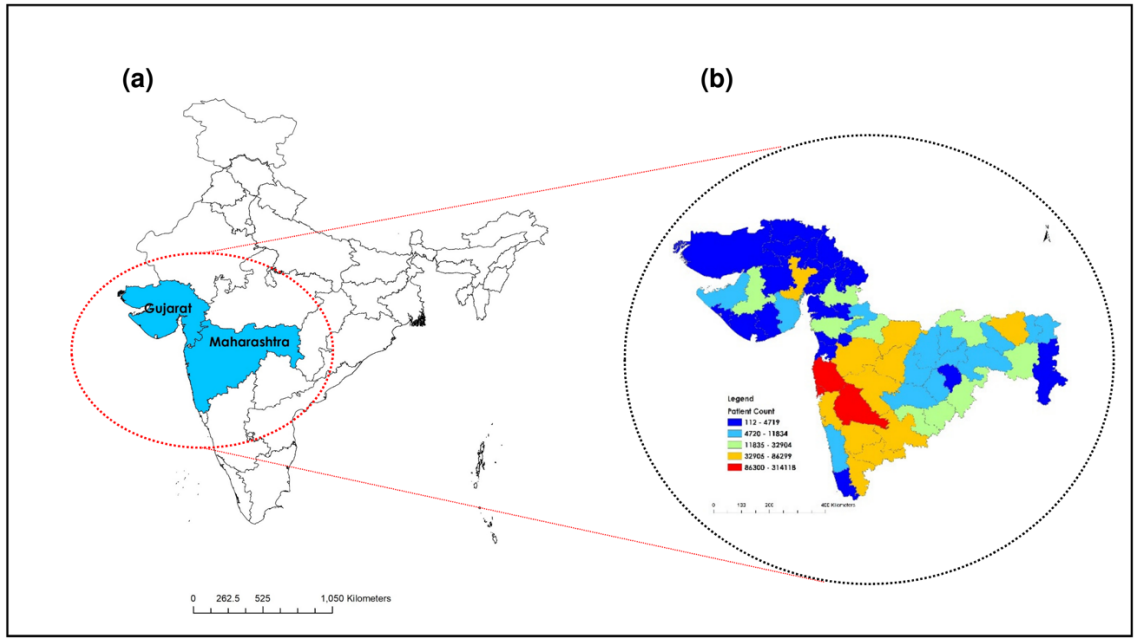

**Figure 1:** (a) Study area of Gujarat and Maharashtra state, India with (b) cumulative SARS-CoV2 Patient count distribution as of October 25, 2020, at the district level.

To identify the identical regions where potential travel corridors can be established, we compute the social and health vulnerability indices using various indicators from 2011 census data (*36*). Specifically, we use the densities of total population, the number of households, fraction of population under six years of age, non-working population, gender ratio, illiterate population, persons with disabilities, and household with electricity aggregated at district scales (*36, 39*). We note that constant growth and migration rates can significantly alter the values of these indicators over the past decade. Here we assume the same relative growth across all the districts in the absence of the latest census data, which is scheduled to be released in the year 2021 (*40*). Further, to calculate the indicators of health vulnerabilities, we use data on dedicated Covid hospitals (DCH), sub-centers, primary healthcare centers (PHC), community healthcare centers (CHC), and sub-divisional hospitals (SDH) sourced from (*41*) and district level SARS-CoV2 patient count (both confirmed and deceased) obtained from (*42*). To construct the transportation network, we use OpenStreetMap to extract the major inter-district roads (*43*).

**Vulnerability Analysis**

It is imperative to categorize the regions that share similar demographics and underlying factors governing social and health vulnerabilities to identify the potential travel corridor. In the proposed framework, we calculate the social and health vulnerability indices for each district (nodes) and health vulnerability index for the road segment (edges) connecting the centroids of neighboring districts. While the associated node vulnerability aids in clustering the similar districts together, the edge vulnerability determines the en-route risk associated due to the prevalent contagious disease. Finally, we integrate the vulnerability indices thus obtained with the transportation network to assess the travel vulnerability between a pair of origin and destination districts, which, in turn, help us identify the possible travel corridors with the least transit risk **(Figure 2)**.

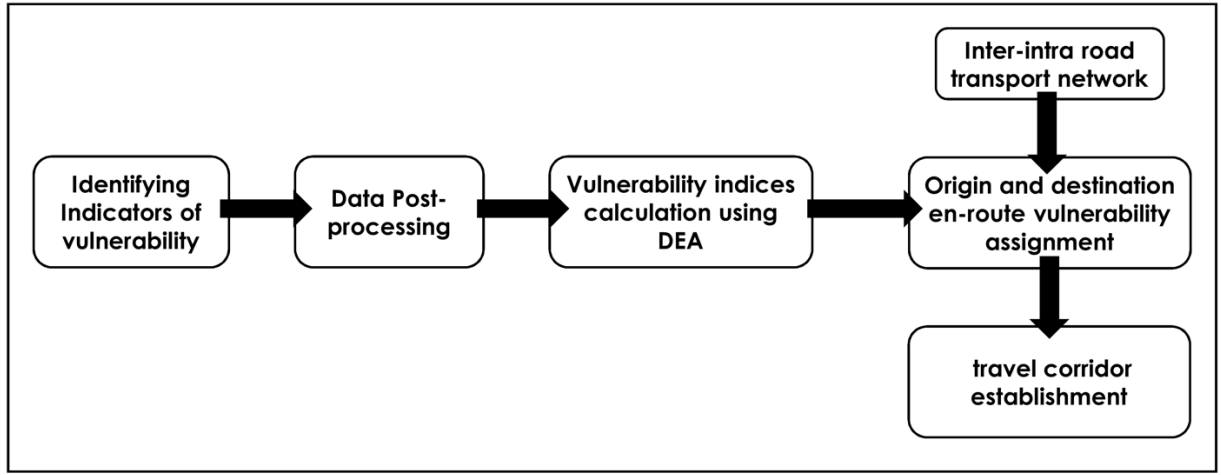

**Figure 2:** Overview of the proposed modeling framework to identify inter-district travel corridors by accounting for social and health vulnerabilities and en-route travel risks during the prevalence of contagious diseases, including SARS-CoV2.

*Variable selection and normalization*

The three approaches typically used by the researchers to select the indicator variables to calculate vulnerability indices include the deductive approach (*44*); the inductive method (*45*); and the hierarchical approach (*31*, *46*). The deductive approach is based on the theoretical understanding of the relationship between parameters. It involves the selection of parameters based on past literature or based on the expert's review (*44*). The hierarchical approach is also based on theoretical understanding, but here, the participatory or expert's opinion is used to select and weigh the parameter. The prime example of this type of approach is the Analytic Hierarchy Process (AHP). On the contrary, in the inductive method, parameters are selected based on the statistical techniques, including Principal Covariate Analysis, to decorate and reduce the dimensionality among the candidate variables for vulnerability assessment. In the present study, we select the same parameters for social vulnerability mapping based on the inductive approach, as identified by (*45*). For the social and health parameters outlined in **Data**, we calculated the density for each parameter by dividing the area of the respective district, and then we use the min-max rescaling to standardize the data (*30*, *45*). Mathematically, the transformation is shown in **Equation 1**:

$$V_i = \frac{Y_i - Y_{min}}{Y_{max} - Y_{min}} \qquad (1)$$

Where:

$V_i$ = Transformed value of parameters used to assess vulnerability scaled between 0 and 1

$Y_{max}$ = Maximum value of the social parameter

$Y_{min}$ = Minimum value of the social parameter

*Data Envelopment Analysis*

The Data Envelopment Analysis (DEA) is a nonparametric mathematical linear programming-based optimization technique originally introduced in (*29*). The DEA calculates the technical efficiency frontier (equivalent to social and health vulnerability) of Decision-Making Units or DMUs (represented by 59 districts in this study). The technique of DEA is widely used in disparate fields, including education (*47*), banking (*48*), transportation (*49*), and the health sector (*50*) for efficiency measurements. In the present study, we use the dual form of Charnes Cooper Rhodes (CCR) model (*29*) described below **(Equations 2-5):**

$Min \quad \theta_k - \epsilon(\sum_{i=1}^{s} S_r^+ + \sum_{i=1}^{m} S_i^-) \qquad (2)$
Subjected to:
$\sum_{j=1}^{n} \lambda_j \, y_{rj} - S_r^+ = y_{rk} \quad \text{where} \quad r = 1,2,\dots,s \qquad (3)$

$\theta_k - x_{ik} \sum_{n}^{j=1} \lambda_j \, x_{ij} - S_i^- = 0 \quad \text{where} \quad r = 1,2,3,\dots m \quad (4)$

$\lambda_j, S_i^-, S_r^+, \epsilon \geq 0 \qquad\qquad\qquad\qquad (5)$

Where:

k= DMU (or district ID)

$\theta_k$= technical efficiency

m = number of inputs

n = number of DMUs

s = number of outputs

$x_{ik}$ = observed magnitude of parameter i for district k

$\lambda_j$ = weight assigned to parameter j

$S_i^-, S_r^+$= slack variables

$y_{rj}$ = observed magnitude of r-type output for entity j

For Social Vulnerability (SV), we considered $y_{rj} = 1$ for all n=59 districts in the absence of any prior data (45). For Health Vulnerability (HV), we considered s=2 where $y_{1j}$ and $y_{2j}$ are the standardized value of confirmed and deceased cases, respectively. The Resultant Vulnerability (RV) of districts is calculated by taking the geometric mean of social and health vulnerabilities (51) **(equation 6):**

$$RV = \sqrt{(SV \times HV)} \quad (6)$$

**Transportation Network Analysis**

To calculate the en-route aggregated risk arising from inter-district travel, we determine inter-district road network data. We connect the centroid of neighboring districts using the shortest path computed using Google Maps. The resulting network thus obtained has district centroid as nodes, and the shortest routes connecting these nodes represent the link. Finally, we calculate the en-route travel risk (TR) to each edge connecting the neighboring districts using **equation 7**:

$$TR_{AB} = \frac{(T_{Ai} \times Hv_a) + (TB_i \times Hv_B)}{T_{AB}} \quad (7)$$

Where

$T_{Ai}$ = time taken from centroid of district A to the boundary of district B

$Hv_a$ = HV of district A

$TB_i$ = time taken from centroid of district B to the boundary of district A

$Hv_B$ = HV of district B

Detailed process-flowchart, along with data used in the present study, is shown in **Figure 3.**

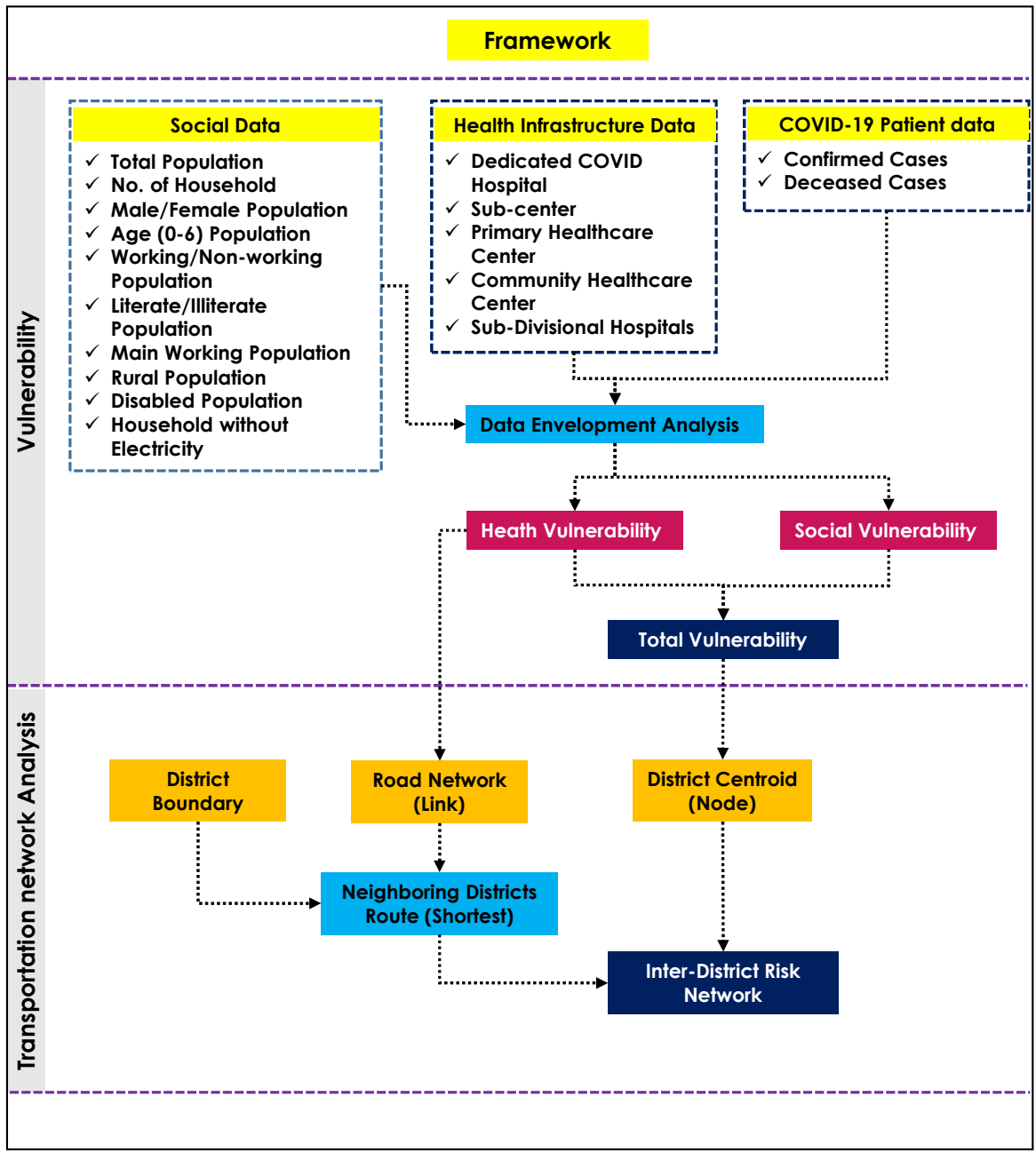

**Figure 3:** Flowchart outlining the datasets and processes involved in health and social Vulnerability assessment to identify low-risk inter-district travel corridors among 59 districts of Gujarat and Maharashtra

**RESULTS AND DISCUSSIONS**

Table 1 provides the summary statistics of parameters used for assessing SV, HV, and TR (in units of quantity per square kilometers). Parameters listed in row 1-12 are used to calculate social vulnerability, whereas the remaining variables are used to calculate health vulnerability and travel risk. These statistics suggest that the overall population density and working population density in Maharashtra are significantly higher than those in Gujarat. However, the mean illiterate population and non-working population density in Maharashtra is also significantly higher than that in Gujarat. A higher population, along with high illiteracy, may have placed Maharashtra in the category of higher vulnerability in comparison to Gujarat. In the case of health infrastructure, the mean density of dedicated Covid hospital and Primary Health Centers (PHCs) are higher in Maharashtra compared to Gujarat. These are the important centers currently working for SARS-CoV2 treatment in India. These Covid hospitals include the Medical Colleges and District hospitals, generally located in the city centers of each district (https://covidindia.org). The PHCs are the centers for collecting test samples for SARS-CoV2 tests. While these facts suggest a better position of Maharashtra to reactively manage the disease, the mean of confirmed and deceased patient count (see Table 1) in Maharashtra is approximately ten times that in Gujarat indicating the prevalent rampant spread of SARS-CoV2. It indicates a vulnerable situation for Maharashtra. Further, these observations were examined in the DEA analysis to identify the vulnerable districts in both states. Note that, officially, there are 36 districts in Maharashtra. However, based on the data availability, the Data for two districts were combined with adjacent districts to observe a total of 34 districts in this Table. Hence, the Greater Mumbai (or Bombay) represents the Mumbai City district and Mumbai Suburban district. Note that Mumbai City and Mumbai Suburban district have integrated transportation system and administration. Similarly, the Thane district in this study represents the Palghar district and Thane districts together. The Palghar district came into existence in 2014, and prior to that, it was part of the Thane district.

Table 1 Descriptive statistics of the density of study variables

| Study variables (Number/km$^2$) | Gujarat *(No. of Districts: 25)* | | Maharashtra *(No. of Districts: 34)* | |
|---|---|---|---|---|
| | Mean | Std. Dev. | Mean | Std. Dev. |
| Population | 411.53 | 228.5 | 729.02 | 2390.7 |
| No. of Households | 83.63 | 48.64 | 160.76 | 534.6 |
| Infant Population | 52.65 | 28.04 | 78.04 | 230.3 |
| Illiterate Population | 127.85 | 61.19 | 168.19 | 448.9 |
| Non-working Population | 241.39 | 140.75 | 421.97 | 1428.7 |
| Female Population | 200.85 | 111.58 | 343.32 | 1099.0 |
| Female Illiterate Population | 76.56 | 36.95 | 94.76 | 242.2 |
| Main working Population | 139.86 | 81.58 | 279.02 | 907.3 |
| Female working Population | 27.45 | 13.63 | 71.13 | 177.5 |
| Rural Population | 254.98 | 141.49 | 136.52 | 83.4 |
| Disabled Population | 7.35 | 5.24 | 19.68 | 65.4 |
| Household without electricity | 8.62 | 5.58 | 14.59 | 13.6 |
| Confirmed patient count | 0.92 | 1.26 | 11.12 | 44.2 |
| Deceased patient count | 0.02 | 0.05 | 0.4 | 1.84 |
| Sub centers ($\times 10^{-3}$) | 50.81 | 31.94 | 38.81 | 47.45 |
| Primary Health Center ($\times 10^{-3}$) | 5.19 | 3.40 | 11.34 | 34.48 |
| Community Health Center ($\times 10^{-3}$) | 2.55 | 1.85 | 1.58 | 1.79 |
| Sub-Divisional Hospitals ($\times 10^{-3}$) | 0.25 | 0.25 | 0.39 | 0.55 |
| Dedicated Covid Hospitals ($\times 10^{-3}$) | 1.05 | 1.29 | 2.42 | 9.57 |

The DEA analysis yields the efficiency scores of DMUs, which, in this study, are the vulnerability scores of the districts considered. Those scores were between 0 and 1, representing the relative vulnerability (risk) of corresponding districts. The vulnerability scores are available for social, health, and a combination of social and health factors. Each factor is classified into five levels representing very high, high, moderate, low, and very low vulnerability. See **Figure 4** for districts with different levels of vulnerability based on social, health, and combinations of social and health factors. The Jenks Natural Breaks Classification technique available in ArcMap® is used to classify the districts. This technique determines the best grouping of vulnerability values based on the user-defined number of classes. It iteratively compares the sums of the squared difference between observed values within each class and class means (*52*). The best classification identifies breaks in the ordered distribution of values that minimizes the within-class sum of squared differences.

As depicted earlier, the observations from **Figure 5(a)** suggests that Maharashtra certainly has higher social vulnerability compared to Gujarat. Among 59 districts from Maharashtra and Gujarat, 25% of districts have high to very high social vulnerability index, out of which, only one district is in Gujarat. Further, the social vulnerabilities in **Figure 4(a)** indicates that the South-West region of Maharashtra has high to very high social vulnerability compared to other districts. Interestingly, the capital of Maharashtra, Mumbai, seems highly vulnerable. Despite being the financial center of India, Mumbai is also home to Asia's largest densely populated slum area called Dharavi. The people in Dharavi are under-privileged with a very low literacy rate. Possibly, for this reason, the social vulnerability of Mumbai is very high. Apart from Mumbai, the DEA analysis indicated that districts very high vulnerability has more illiterate population, non-working population and high infant population. Further, the districts with very low to low social vulnerability have higher number of working female population. In Gujarat, The Kachchh district is identified as the region of high social vulnerability owing to high illiteracy rate, larger rural population with marginal workers and higher density of households without electricity. Other districts in Gujarat are very low to moderately vulnerable.

While a well-connected transportation network is conducive to the economic prosperity of the region, it can open the gateways for entry of new cases during the prevalence of a pandemic. However, the health infrastructure could be insufficient to deal with such a pandemic. In the context of health vulnerability for Gujarat: Patan, Porbandar, Vadodara and Narmada have shown very high vulnerability as the health infrastructure available to tackle the SARS-COV2 is very low compared to the patient cases in the respective districts. Whereas Ahmedabad and Surat being the financial hubs of Gujarat have shown high vulnerability due to the high SARS-COV2 patient cases (**Figure 4(b)**). For Maharashtra, Mumbai has one of the busiest single-runway International airports in the world and is well connected by roads and railways with other parts of the country. Any individual with Covid-19 infection entering Mumbai can contribute to the rapid spreading of disease in Mumbai as well as in the nearby districts of Maharashtra. Districts with good connectivity with Mumbai have reported a higher number of cases compared to other districts. Mumbai is located on the West coast of Maharashtra. Hence, the Westen region of Maharashtra state, as shown in **Figure 4(b)**, has a high patient count and more vulnerable districts compared to its Eastern region and Gujarat state. It includes prominent districts such as Thane, Pune, Nasik, Aurangabad, and Sangali. In the Eastern Part of Maharashtra, the Nagpur district showed high health vulnerability. Note that it is well connected with Mumbai and other prominent cities in India through various transport modes (air, rail, and road). Being at the center of India, Nagpur airport is strategically developing into a Multi-modal International Cargo Hub for goods transportation. Since it is not fully operational to date, the neighboring districts relatively have a lesser vulnerability. Overall, **Figure 4(c)** suggests that the total vulnerability was very high in Maharashtra compared to Gujarat. It seems logically correct since the number of SARS-CoV2 patients in Maharashtra is almost ten times that of Gujarat.

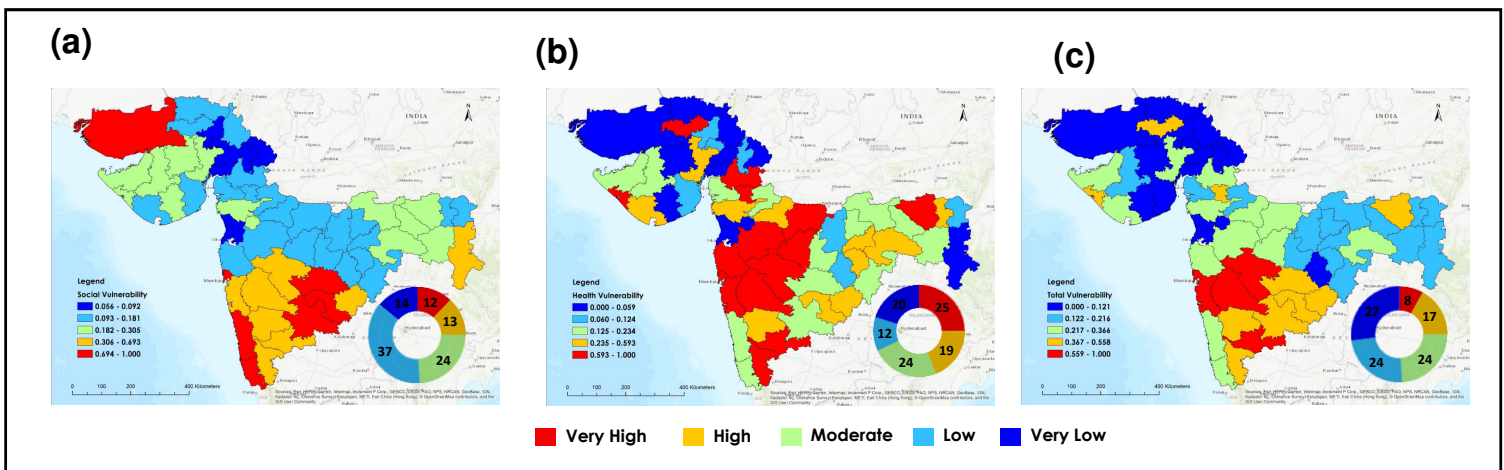

**Figure 4:** DEA analysis-based assessment of (a) Social Vulnerability, (b) Health Vulnerability, and (c) Total Vulnerability (geometric mean of social and health vulnerability). Pie-chart in the inset of each figure shows the percentage of districts belonging to each class.

As mentioned earlier, each district is represented as a node at the centroid for the network analysis. The shortest route (edge) between neighboring district centroids was obtained from the Google maps$^{©}$. The total vulnerability is assigned to the centroids/nodes of the district. However, for the links, a weighted health vulnerability is assigned to identify risk routes. It is based on the hypothesis that individuals living/staying in a district would be vulnerable to the surrounding social characteristics and their health vulnerability will be a function of SARS-CoV2 cases and health infrastructures of the district. However, while traveling, they are only vulnerable to the en-route SARS-CoV2 risk. The weights of such links are calculated using **equation 7.** The nodes and links for Maharashtra and Gujarat districts are color-coded as per the RV classification (**Figure 5a**). The respective color code for nodes represents the district vulnerability, while links provide the neighboring district travel vulnerability. **Figure 5 (a)** illustrates the weighted risk on the links of neighboring districts, Figure 5(b) shows 15x15 en-route travel risk matrix sampled from the 59x59 matrix for better readability. Based on the network, associated matrix, and calculations illustrated in **Figure 6;** a user can decide whether to travel on a particular link and choose a comparatively low-risk route for traveling from source to neighboring destination district. The weighted travel network thus obtained can help us understand the overall travel risk. For example, Vadodara and Narmada, the two districts of Gujarat, have low and high resultant vulnerabilities, respectively. However, the en-route travel risk is obtained as very high owing to the greater health vulnerability of the Vadodara district, which in turn can be explained by a large number of reported cases and larger centroid to boundary distance that commuter will spend in Vadodara than Narmada.

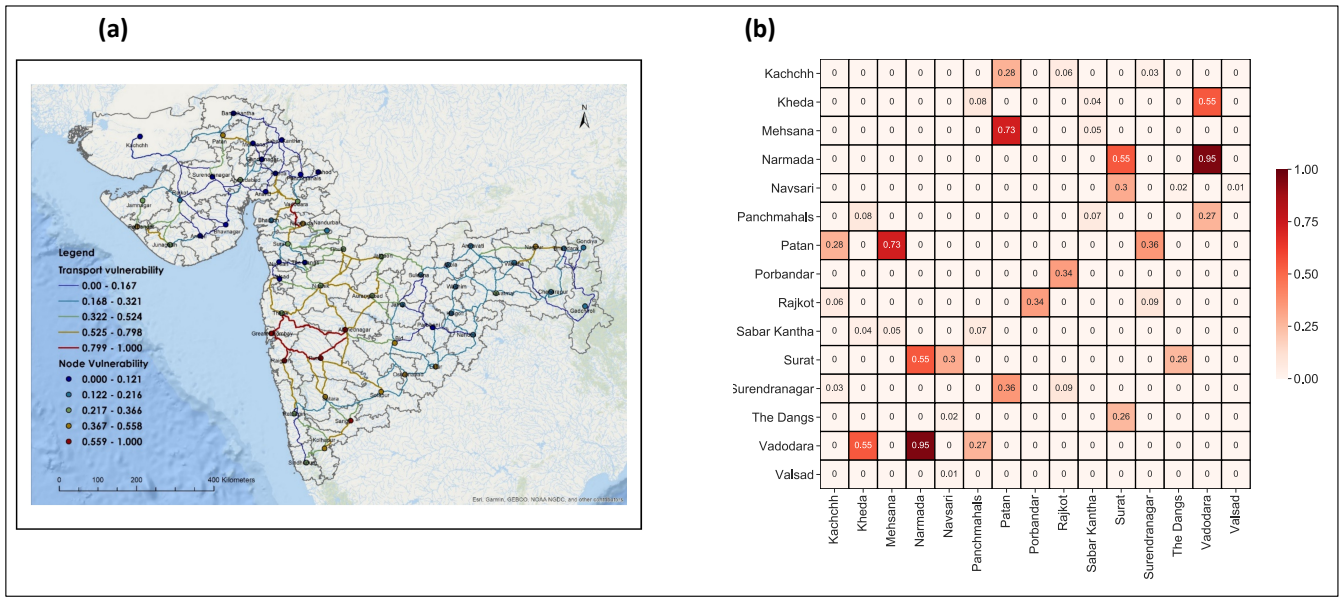

**Figure 5: (a)** Interdistrict risk network, where nodes represent the Total Vulnerability (TV) of district and road network (Links) depicting the risk associated with inter-district travel. **(b)** En-route travel risk matrix for 15 districts.

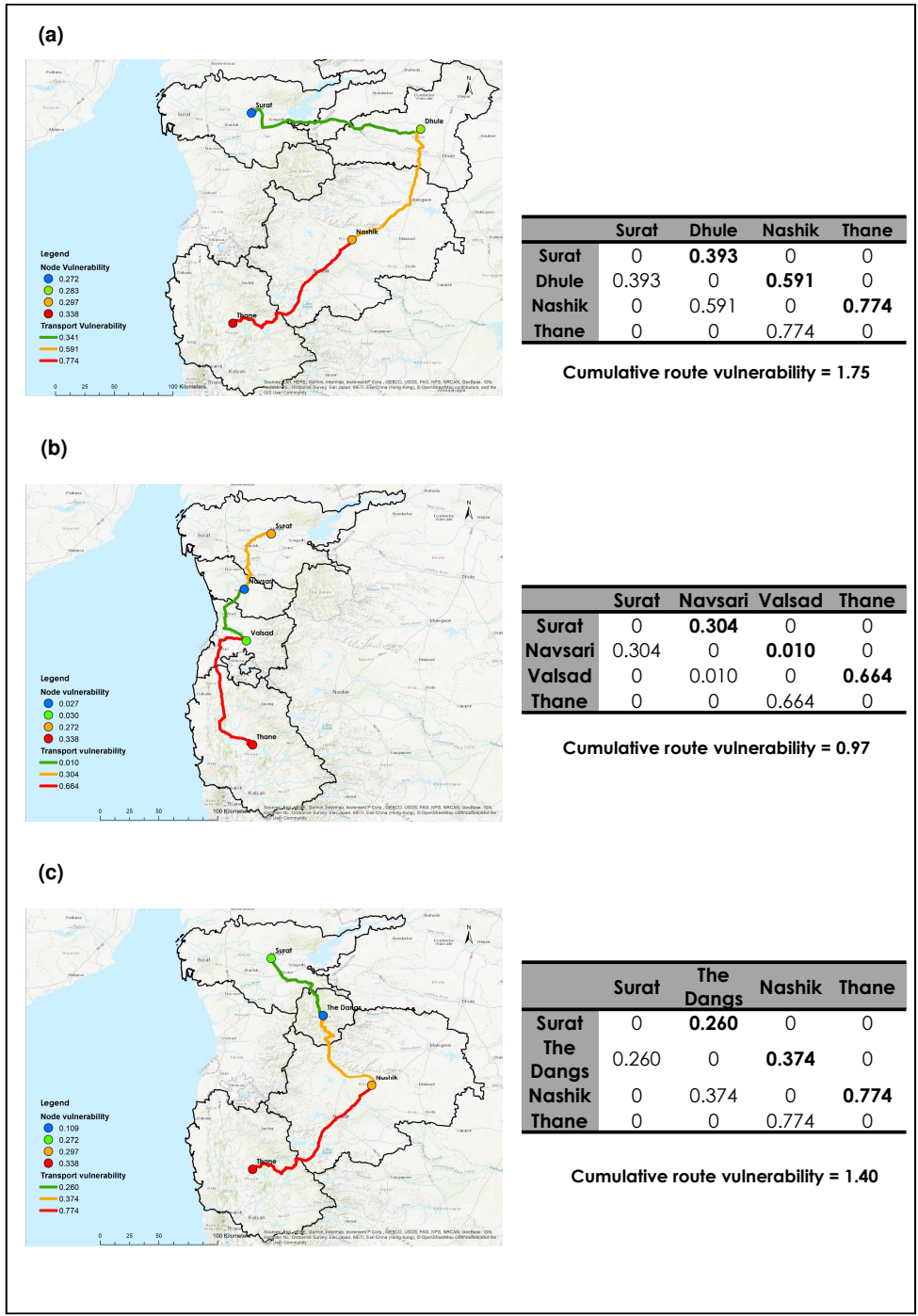

**Figure 6:** Illustrative example showing cumulative en-route travel risk between origin (Surat)-destination (Thane). The cumulative vulnerability is calculated by summation of the cells above (or below) the diagonal of the en-route Travel Risk (TR) matrix (a) Route via Surat- Dhule- Nashik- thane results in the cumulative travel risk of 1.75. Similarly, the route via (b) Surat- Navsari- Valsad- Thane

yields the cumulative vulnerability of 0.97, and the route via (c) Surat- The Dangs- Nashik- Thane results in the cumulative TR of 1.4, making route (b) the safest choice among all three.

We note that in a real-world network, including the one developed here, multiple alternate routing options (without considering the shortest distance) are feasible between a pair of districts. Hence, it is critical to know how many n-walk routes exist between an origin-destination pair. To address this, we calculate the adjacency matrix $A_{ij}$ of the network shown in figure **7a**. Element $a_{ij}$ of the matrix $A_{ij}$ would be equal to 1 if the district $j$ is connected to district $i$, and 0 otherwise (**Figure 7a**). The n-th power of $A_{ij}$ thus, obtained would yield the total number of n-length walks between a pair of districts along which we would perform the travel risk calculations to identify the path with least en-route risk (**Figure 7b**) using procedures outlined in **Figure 6.**

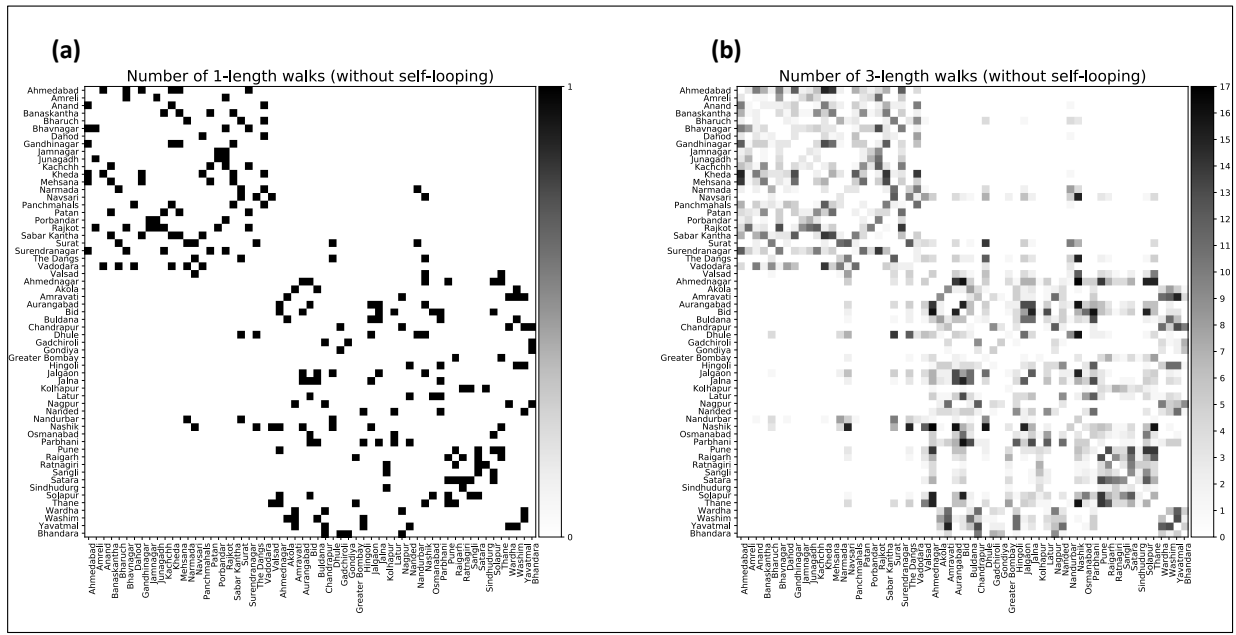

**Figure 7: (a)** Adjacency matrix, $A_{ij}$ of the network shown in (a) indicates the presence of 1 length walk. **(b)** 3[rd] power of $A_{ij}$ indicates the presence of 3 length walks. Diagonal of the matrices are made zero to avoid self-loops.

.
**CONCLUSIONS**

Lockdowns to curb the spread of Novel Coronavirus (SARS-CoV2) brought transportation systems to stand still, severely impacting the economic, trade, and tourism activities worldwide. As nations and states plan to open up their borders despite the prevalent pandemic, there is a greater need to make informed decisions to keep the virus spread in check. In this paper, we develop a social and health vulnerability informed quantitative framework to identify the potential low-risk travel corridors for interdistrict travel. The risk of contagion depends on the disease prevalence in the community at a given time. Therefore, in addition to assessing the associated risks at origins and destinations, travel corridors should account for en-route travel risks to identify the safest shortest routes and detours. Even though we demonstrate the application of the proposed model for establishing travel bubbles within and across Gujarat and Maharashtra, the proposed plan can be applied to other regions for domestic and internation travel planning with appropriate contextualization. Key challenges that need to be addressed to scale-up and operationalize the proposed strategy include the availability of homogeneous socio-economic vulnerability data at local scales and continuous data streams associated with regions' real-time contagion spread. We note specific caveats related to the demonstration of our proposed framework. While we have assumed the uniform travel speed throughout the network, real-time mobility patterns accounting for more accurate descriptions of congestions can alter the en-route risk factors dynamically. Further, we have used the socio-economic indicators from the 2011 census given the decadal frequency of census surveys in India. However, these states have witnessed substantial worker migration from various regions in the past decade (*53*). Hence, accounting for these changes can change the social vulnerability classification of highly industrialized districts. Specific clinical interventions including vaccines or enforcing/lifting lockdown measures lie outside the scope of public transportation managers, but measures discussed in this work can help them in risk-informed decision making, issue travel advisories, and establish travel corridors that pose the least risk to the public.

## AUTHOR CONTRIBUTIONS

The authors confirm contribution to the paper as follows: study conception and design: U. Bhatia and A. Maji; data collection: R Dave, T. Choudhari; analysis and interpretation of results: R Dave, T. Choudhari, U. Bhatia, and A. Maji; draft manuscript preparation: R Dave, T. Choudhari, U. Bhatia and A. Maji. R. Dave and T. Choudhari contributed equally. All authors reviewed the results and approved the final version of the manuscript.